\documentclass[aps,prl,twocolumn,superscriptaddress,showpacs,]{revtex4-1}

\usepackage{lineno}
\usepackage{color}
\usepackage{amsmath}
\usepackage{amssymb}
\usepackage{graphics}
\usepackage{graphicx}
\usepackage{epstopdf}
\usepackage{dcolumn}
\usepackage{bm}
\usepackage{xcolor}
\usepackage{hyperref}
\hypersetup{hidelinks}

\newcommand{\BE}{{\bf E}}
\newcommand{\BW}{{\bf W}}
\newcommand{\Be}{{\bf e}}
\newcommand{\Br}{{\bf r}}
\newcommand{\BR}{{\bf R}}

\newcommand{\BA}{{\bf A}}

\newcommand{\BF}{{\bf F}}

\newcommand{\mtW}{{\mathcal{W}}}
\newcommand{\BmtW}{{\bm{\mathcal{W}}}}
\newcommand{\Bkappa}{\bm{\kappa}}
\newcommand{\Bdelta}{\bm{\delta}}

\begin{document}
\title{Phase-space nonseparability, partial coherence, and optical beam shifts}
\author{Yahong Chen}
\email[]{yahongchen@suda.edu.cn}
\affiliation{School of Physical Science and Technology, Soochow University, Suzhou 215006, China}
\affiliation{Suzhou Key Laboratory of Intelligent Photoelectric Perception, Soochow University, Suzhou 215006, China}
\author{Sergey A. Ponomarenko}
\email[]{serpo@dal.ca}
\affiliation{Department of Electrical and Computer Engineering, Dalhousie University, Halifax, Nova Scotia, B3J 2X4, Canada}
\affiliation{Department of Physics and Atmospheric Science, Dalhousie University, Halifax, Nova Scotia, B3H 4R2, Canada}
\begin{abstract}
As a paraxial wave packet is reflected or refracted from a planar interface separating two material media, it experiences spatial and angular shifts of its center position with respect to predictions of the geometrical ray picture. These in-plane and out-of-plane beam shifts are known as Goos-H\"{a}nchen and Imbert-Fedorov shifts, respectively. We discover a universal link between the phase-space nonseparability of an incident wave packet of any degree of spatial coherence and the reflected beam shifts. We unveil coherence Goos-H\"{an}chen and coherence Hall effects, absent in the fully coherent limit. While the former effect can trigger a pronounced enhancement of the spatial Goos-H\"{an}chen shift, the latter enables control of the spatial Imbert-Fedorov shift, from complete cancellation at a certain incidence angle to dramatic enhancement of the shift to giant magnitudes for nearly incoherent incident wave packets. Our results are equally applicable to optical, X-ray, neutron, as well as matter waves, and they showcase novel phenomena in wave-matter interactions.
\end{abstract}
\maketitle

The spatial Goos-H\"{an}chen (GH)~\cite{goos1947neuer} and Imbert-Fedorov (IF)~\cite{fedorov1955k, Imbert72} shifts of optical beams reflected from a flat interface separating two dielectric media, which are among the most striking manifestations of the wave nature of light, were originally discovered in the context of total internal reflection. The discovery was followed by the disclosure of angular GH and IF shifts in partial reflection \cite{Ra73, Merano09} and the recognition that similar wave phenomena occur in reflection/refraction of neutrons~\cite{ignatovich2004neutron,de2010observation}, X-rays~\cite{tamasaku2002goos}, electrons~\cite{beenakker2009quantum,wu2011valley}, as well as spin~\cite{dadoenkova2012huge} and matter waves~\cite{huang2008goos,mckay2025observation} from spatial or spatio-temporal~\cite{ponomarenko2022goos} interfaces of two material media. In addition, GH and IF shifts, the latter is associated with the so-called spin-Hall effect~\cite{Onoda04,Bliokh06,Hosten08,bliokh2013goos,Kim23}, found numerous applications to temperature~\cite{zhou2021temperature} and humidity~\cite{wang2015optical} sensing, diﬀerential microscopy~\cite{Zhu19Generalized,Wang22Photonic,Liu24PRL}, as well as the reflectometry of structured surfaces, including resonant structures~\cite{wu2019giant,wu2021observation} and layered magnetic materials~\cite{frank2014goos}.

Although GH and IF shifts of conventional Gaussian beams have been extensively researched, spatial and angular shifts of structured wave packets have sparked growing interest. In particular, the orbital angular momentum (OAM) content of vortex-endowed light beams has been shown to affect the magnitude of the shifts~\cite{fedoseyev2001spin,dasgupta2006experimental,okuda2008significant,bliokh2009goos,bekshaev2009oblique,merano2010orbital,Dennis12,Loffler12,Barros24}, as has the quantum entanglement of spatial and polarization degrees of freedom of incident wave packets~\cite{le2025entangled}. At the same time, there has been a long-standing controversy vis-\`{a}-vis the effect of spatial coherence of the incident light source on the spatial GH shift. While some theoretical~\cite{simon1989nonspecular,aiello2011role,tang2025effect} and experimental~\cite{loffler2012spatial,merano2012observation} studies found no link between the source coherence and the magnitude of the spatial GH shift, others came to diametrically opposite conclusions~\cite{wang2008influence,wang2013goos}. 

Here, we advance a general phase-space theory of spatial and angular shifts in reflection of paraxial wave packets from a material interface employing the Wigner distribution function (WDF). Our theory reveals a fundamental connection between the phase-space nonseparability of the incident and optical shifts of the reflected wave packets and it clarifies the role of spatial coherence of the incident wave packets in their interaction with material interfaces. Specifically, using a twisted Gaussian Schell-model source \cite{Simon93,sundar1993twisted,ponomarenko2021twist} as a representative example of a structured partially coherent, phase-space nonseparable wave packet, we disclose coherence GH and coherence Hall effects that do not exist in the fully coherent limit. While the former effect causes a pronounced enhancement of the spatial GH shift, the latter enables control of the spatial IF shift, from complete cancellation at a certain incidence angle to dramatic enhancement of the shift for nearly incoherent incident wave packets. Although we put our theory in the optical context, we stress that the WDF enables a universal description of beam shifts, from classical to quantum waves of any physical nature, from optical to matter waves. 

We start by considering a statistical ensemble of electric fields $\{\BE_i\}$ of paraxial electromagnetic wave packets (beams) incident at the interface $z=0$ that separates two transparent non-magnetic media. We can express the cross-spectral density matrix of the ensemble at a pair of points $\Br_1$ and $\Br_2$ at the interface as~\cite{mandel1995optical} $\BW_i (\Br_1,\Br_2,\omega)=\langle \BE_i^*(\Br_1,\omega)\BE_i^{\mathrm{T}}(\Br_2,\omega)\rangle$,
where the asterisk, T and the angle brackets denote complex conjugation, transposition and ensemble averaging, respectively; the subscript ``$i$" refers to an incident beam. Assuming our beams to be quasi-monochromatic, we can drop the frequency dependence hereafter. Next, we introduce a WDF matrix of the incident beam ensemble as
\begin{equation}\label{Wigmat-def}
    \bm{\mtW}_i(\Bkappa,\BR)=\int d\Br\,\BW_i(\BR-\Br/2,\BR+\Br/2)e^{-\mathrm{i}\Bkappa\cdot\Br},
\end{equation}
where $\BR=(\Br_1+\Br_2)/2$ and $\Br=\Br_2-\Br_1$. We note in passing that the WDF of any quantum density operator is defined by replacing the cross-spectral density matrix with the corresponding matrix element of the density operator~\cite{schleich2015quantum}; since we utilize only second-order correlation functions, classical and quantum descriptions should yield the same results. 

\begin{figure}[t]
\centering
\includegraphics[width = 0.9 \linewidth]{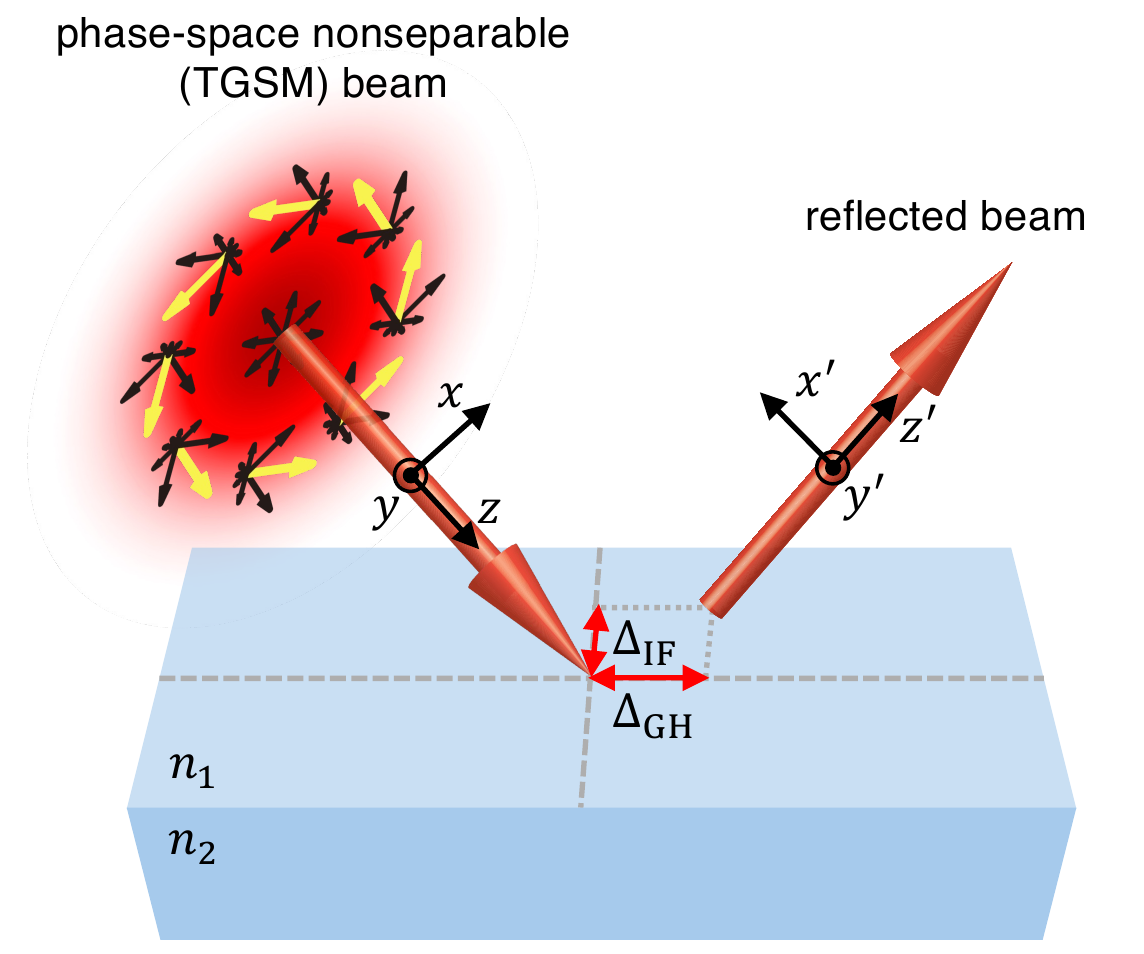}
\caption{Illustrating the reflection of a phase-space nonseparable wave packet, such as a twisted Gaussian Schell-model (TGSM) beam at a dielectric interface. The arrows in the cross-section of the incident beam represent the distribution of transverse wavevectors at different spatial positions. The central transverse wavevectors at each spatial point are highlighted in yellow; in the TGSM case, the yellow arrows exhibit an azimuthal distribution with handedness determined by the sign of the twist factor $u$ of the TGSM beam. This spatially inhomogeneous transverse wavevector distribution manifests the phase-space nonseparability of the beam, which, in turn, gives rise to the coherence-controlled spatial GH ($\Delta_\mathrm{GH}$) and IF ($\Delta_\mathrm{IF}$) shifts upon reflection.}
\label{fig1}
\end{figure}

We can write the WDF matrix $\BmtW_i$ of any ensemble of uniformly polarized incident wave packets as a product of the polarization matrix $\BA$ and scalar WDF $\mtW_i$ as
\begin{equation}\label{WDM}
    \BmtW_i (\Bkappa,\BR)=\BA\mtW_i(\Bkappa,\BR),
\end{equation}
where 
\begin{equation}\label{A}
    \BA=\begin{pmatrix}
    |a_x|^2 & a_x^* a_y \\ a_xa_y^* & |a_y|^2
        \end{pmatrix}.
\end{equation}
Here $a_{x}$ and $a_y$ are the amplitudes of in-plane and out-of-plane polarization components, respectively; $x$- and $y$-axes are attached to an incident beam (see Fig.~\ref{fig1}). Further, the scalar WDF is defined in terms of the scalar cross-spectral density by the equation strictly analogous to Eq.~(\ref{Wigmat-def})~\cite{SM}.

We are now in a position to define the centroid position vector of the beam as
    \begin{equation}\label{r0-def}
        \langle\langle\BR_{c}(z')\rangle\rangle = \frac{\int d\Bkappa\int d\BR'\,\BR'\mathrm{Tr}[\bm{\mtW}_r(\Bkappa,\BR',z')]}{\int d\Bkappa\int d\BR'\,\mathrm{Tr}[\bm{\mtW}_r(\Bkappa,\BR',z')]},
    \end{equation}
where the subscript ``$r$" refers to the quantities and the prime to the coordinates associated with the reflected beam (Fig.~\ref{fig1}). Next, we can determine the centroid of the reflected beam at any distance $z'$ away from the interface in the Wigner picture (see Sec.~S2 of the Supplemental Material \cite{SM}), 
\begin{equation}\label{Rc1}
\langle\langle\BR_{c}\rangle\rangle=\bm{\Delta}+\bm{\delta}_{R}+\bm{\delta}_{\Theta},
\end{equation}
where $\bm{\Delta}$, given explicitly in Eq.~(\ref{Del}) of the End Matter, is independent of the phase-space structure of the incident polarized beam. At the same time, the other two terms on the right-hand side of Eq.~(\ref{Rc1}) describe spatial ($\Bdelta_R$) and angular ($\Bdelta_{\Theta}$) shifts that depend on the WDF matrix $\BmtW_i$ of the incident beam. These shifts read
\begin{equation}\label{DelR}
\begin{split}
\Bdelta_R = \; & 2\left[\overline{\BR'\kappa_y'}\,\frac{|r_s|^2\mathrm{Re}(Y_sa_xa_y^*)-|r_p|^2\mathrm{Re}(Y_pa_x^*a_y)}{|a_x|^2|r_p|^2+|a_y|^2|r_s|^2} \right. \\
 &\left. + \; \overline{\BR'\kappa_x'}\,\frac{|a_x|^2|r_p|^2 \mathrm{Re}X_p+|a_y|^2|r_s|^2 \mathrm{Re}X_s}{|a_x|^2|r_p|^2+|a_y|^2|r_s|^2}\right],
\end{split}
\end{equation}
and
\begin{equation}\label{Delthta}
\begin{split}
    \Bdelta_{\Theta}\!=& \; \frac{2z'}{k}\left[\overline{\Bkappa'\kappa_y'}\,\frac{|r_s|^2\mathrm{Re}(Y_sa_xa_y^*)-|r_p|^2\mathrm{Re}(Y_pa_x^*a_y)}{|a_x|^2|r_p|^2+|a_y|^2|r_s|^2} \right. \\
 &\left. + \; \overline{\Bkappa'\kappa_x'}\,\frac{|a_x|^2|r_p|^2 \mathrm{Re}X_p+|a_y|^2|r_s|^2 \mathrm{Re}X_s}{|a_x|^2|r_p|^2+|a_y|^2|r_s|^2}\right].
\end{split}
\end{equation}

\noindent Here $k=n_1\omega/c$, $n_1$ being the refractive index of the upper half-space (see Fig.~\ref{fig1});  $r_{s,p}$ are (complex) reflection coefficients of $s$- and $p$-polarizations and $X_{s,p}$ as well as $Y_{s,p}$ are the corresponding shifts of fully coherent, phase-space separable, polarized beams [see Eq.~(\ref{XY}) of the End Matter]. Further, we introduce a phase-space average of any, in general, vector function $\BF(\Bkappa,\BR)$ with respect to the WDF matrix of the incident beam ensemble as
\begin{equation}\label{ave-def}
    \overline{\BF(\Bkappa,\BR)}=\frac{\int d\BR\int d\Bkappa \BF(\Bkappa,\BR)\mathrm{Tr}[\bm{\mtW}_i(\Bkappa,\BR)] }{\int d\BR\int d\Bkappa\mathrm{Tr}[\bm{\mtW}_i(\Bkappa,\BR)]}.
\end{equation}
We note that the marginals of the WDF matrix provide the angular spread and spatial intensity distribution of the incident beam:
\begin{subequations}\label{IJ-def}
    \begin{eqnarray}
        J(\Bkappa)=\int d\BR \mathrm{Tr}[\bm{\mtW}_i(\Bkappa,\BR)] , \\
        I(\BR)=\int \frac{d\Bkappa}{(2\pi)^2}\mathrm{Tr}[\bm{\mtW}_i(\Bkappa,\BR)] .
    \end{eqnarray}
\end{subequations}
We can readily infer from Eqs.~(\ref{ave-def}) and (\ref{IJ-def}) that for any paraxial beam, centered on its axis, we must have
\begin{gather}\label{Rkap=0}
    \overline\BR=\frac{\int d\BR \BR I(\BR)}{\int d\BR I(\BR)}=0; \quad
        \overline\Bkappa=\frac{\int d\Bkappa \Bkappa J(\Bkappa)}{\int d\Bkappa J(\Bkappa)}=0.
\end{gather}
In view of Eq.~(\ref{WDM}), the phase-space separability of any fully polarized source can be quantified in terms the scalar WDF $\mtW_i$ as
\begin{equation}\label{WDF-sep}
    \mtW_i(\Bkappa,\BR)=f(\Bkappa)g(\BR),
\end{equation}
where $f$ and $g$ are any scalar functions. It then follows at once from Eqs.~(\ref{DelR}), ~(\ref{Rkap=0}) and (\ref{WDF-sep}) that spatial and momentum averages in Eq.~(\ref{DelR}) factorize, leading to $\Bdelta_R =0$ for any phase-space separable incident beam. The phase-space nonseparability of a source leads to an inhomogeneous momentum distribution of the incident wave packet (see, e.g., in Fig.~\ref{fig1}), which, in turn, affects the spatial GH and IF shifts profoundly through the $\Bdelta_R$ cross-term. At the same time, we can infer from Eq.~(\ref{Delthta}) that the angular shifts are unaffected by the phase-space nonseparability of the source. For this reason, we focus on spatial GH and IF shifts henceforth. 

We now show that in the low-coherence limit, a phase-space nonseparable incident beam can acquire large spatial GH and IF shifts in partial reflection from the interface of transparent media. To make our findings concrete, we consider an incident twisted Gaussian Schell-model (TGSM) beam, which serves as a representative example of a partially coherent phase-space nonseparable wave packet. The TGSM beams were theoretically introduced in pioneering works~\cite{Simon93,sundar1993twisted}, experimentally realized~\cite{Ari94,Wang19,Wang25} and have played a role in random soliton theory~\cite{PSA01}, spin-orbit interactions with random light~\cite{Yahong25}, optical information transfer through turbulence~\cite{Cai06,Wang12}, optical imaging~\cite{Olga09,Olga12} and spontaneous parametric down-conversion engineering \cite{Hutter20}.


In Sec. S2 of the Supplemental Material~\cite{SM}, we derive a closed form expression for the centroid position of a reflected TGSM beam as
\begin{equation}\label{Rc}
    \langle\langle \BR_c (z')\rangle\rangle=\bm{\Delta} - \frac{2t}{\xi_c^2}\bm{\Lambda}+\frac{2z'}{z_R\xi_c^2}\left(1+\frac{\xi_c^2}{4}+\frac{t^2}{\xi_c^2}\right)\bm{\Theta}.
\end{equation}
Here $z_R=k\sigma_I^2$ is a Rayleigh range associated with the width of the source; $t$ is a dimensionless twist parameter, such that $-1\leq t=u\sigma_c^2\leq 1$, and $\xi_c=\sigma_c/\sigma_I$ is a dimensionless coherence parameter, related to dimensional rms beam and coherence widths at the source, $\sigma_I$ and $\sigma_c$, as well as the twist parameter $u$ [see Eqs.~(\ref{WDF-A}) through~(\ref{u}) of the End Matter]. In addition, the coefficients $\bm{\Lambda}$ and $\bm{\Theta}$ are also explicitly given in Eqs.~(\ref{Lda}) and (\ref{Tta}) of the End Matter.

The analysis of Eqs.~(\ref{Rc}) and~(\ref{Del}) through (\ref{XY}) reveals that the second term on the right-hand side of Eq.~(\ref{Rc}) can dominate spatial GH and IF shifts in the low coherence limit, $\xi_c\rightarrow 0$, yielding giant shifts. The sign of either shift is governed by the twist parameter $t$. Since the twist phase vanishes in the fully coherent limit~\cite{Simon93} (see also End Matter), we refer to the corresponding effect as a \textit{coherence GH shift}. Note that all previous studies of GH shifts with partially coherent wave packets~\cite{simon1989nonspecular,aiello2011role,tang2025effect,loffler2012spatial,merano2012observation,wang2008influence,wang2013goos} were concerned with Gaussian Schell-model beams which are phase-space separable, and hence do not exhibit this novel effect. In the limit $t=0$, the coherence GH shift vanishes and our results are in agreement with~\cite{simon1989nonspecular,aiello2011role,tang2025effect,loffler2012spatial,merano2012observation}, but contrary to~\cite{wang2008influence,wang2013goos}. 

We can explain the spatial GH and IF shift enhancement of nearly incoherent incident beams by noticing that in the low-coherence limit, the WDF of a TGSM source has a wide momentum distribution, see Eqs.~(\ref{Wig-TGSM}) and~(\ref{sigef}) of the End Matter, which, in turn, induces a large spatial shift due to phase-space non-separability. Alternatively, the phase-space nonseparability of a TGSM beam source is manifest by the circulation of the wavevector field around the beam axis, see Fig.~\ref{fig1}, with the OAM of $-2t\hbar/\xi_c^2$ per photon~\cite{gbur16}. Therefore, the coherence GH shift of the TGSM beam is similar to that due the OAM of a fully coherent vortex beam. However, unlike the latter, the TGSM beam has a Gaussian intensity profile with no axial singularity. Therefore, its OAM is akin to that of a solid rotor, which scales with the rotor size, rather than being fixed by the topological charge of the beam~\cite{gbur16}. As a result, the TGSM OAM can reach extremely large magnitudes for nearly incoherent, wide beams, causing huge GH and IF shifts. In contrast, the enhancement of the spatial shifts due to a coherent vortex is limited in practice by our inability to generate noise-free vortices with higher-order topological charges \cite{Wang18recent}.

\begin{figure}[b]
\centering
\includegraphics[width= \linewidth]{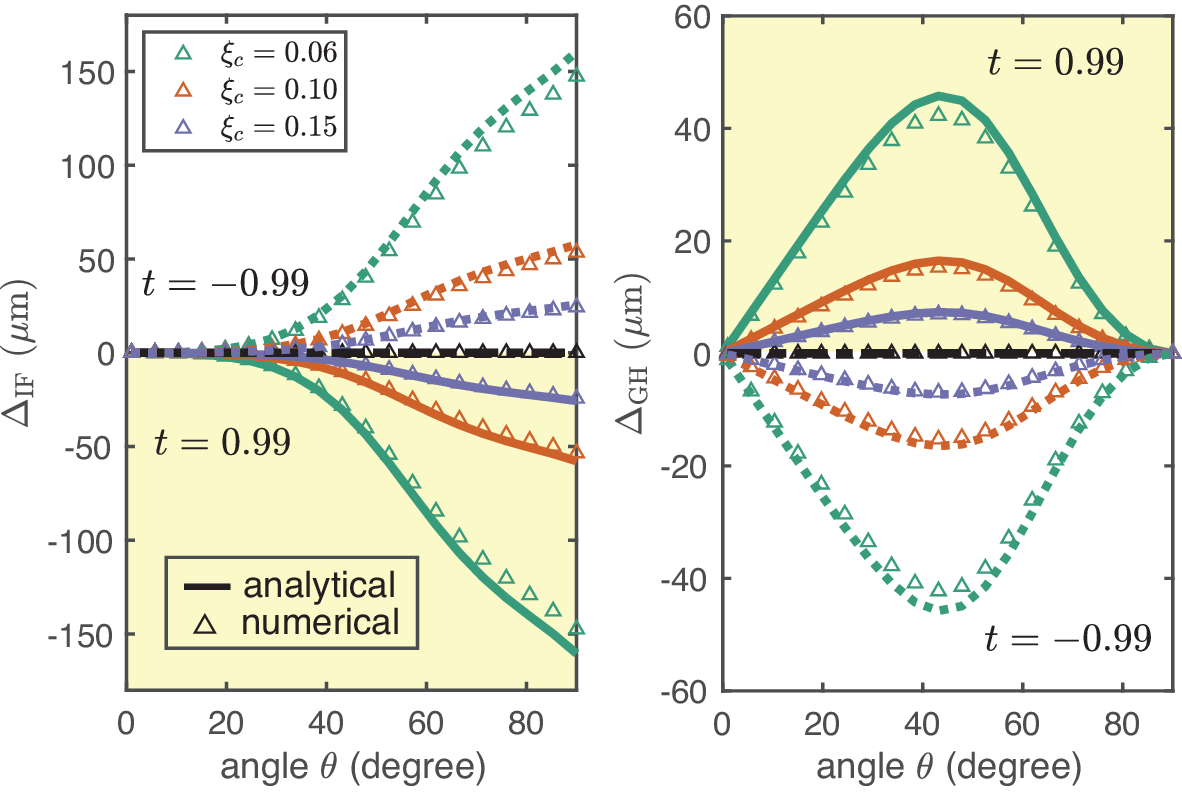}
\caption{Analytical (curves) and numerical (triangles) results for the spatial IF and GH shifts of a linearly polarized TGSM beam under partial reflection and variable $t$ and $\xi_c$. The simulation parameters are: $n_1 = 1$, $n_2 = 1.515$, $a_x = a_y = 1$, $\lambda_0 = 633$~nm, $\sigma_I = 1$~mm. The solid, dotted, and dash-dotted curves correspond  to $t = 0.99$, $t = -0.99$, and $t = 0$.}
\label{fig2}
\end{figure}

We now validate our analytical results with numerical simulations, employing the angular spectrum decomposition and coherent-mode representation of the cross-spectral density matrix of a TGSM beam ensemble~\cite{SM}. We first consider the case of partial reflection, where the refractive indices of the two media are $n_1 = 1$ (air) and $n_2 = 1.515$ (glass). In addition, we consider a linearly polarized incident TGSM beam with $a_x = a_y = 1$, at wavelength $\lambda_0 = 633$~ nm, and a beam waist of $\sigma_I = 1$~mm. In Fig.~\ref{fig2} we show the dependence of spatial GH and IF shifts on the incidence angle $\theta$ for the variable $\xi_c$ marked with the corresponding colors. The solid, dotted, and dash-dotted curves correspond to our analytical results obtained from Eq.~(\ref{Rc}) for $t = 0.99$, $t = -0.99$, and $t = 0$, respectively, and our numerical results are marked with colored triangles. We observe marked enhancement of GH and IF shifts for large $t$ and small $\xi_c$, manifesting very good agreement between the theory and simulations. Instructively, we find that even for linearly polarized incident beams, which carry no spin angular momentum, the twist phase leads to a transverse IF shift upon reflection. This phenomenon resembles the optical spin Hall effect, in which the spin of the incident beam induces a transverse IF shift. Here, however, the shift is induced by the beam coherence rather than spin, and we dub this the \textit{coherence Hall effect} in reflection. Importantly, while spin-induced IF shifts are typically shorter than the wavelength, the coherence Hall effect produces coherence controlled shifts that attain values as large as $240$ wavelengths for $\xi_c = 0.06$ in our simulations (within the paraxial region).

\begin{figure}[b]
\centering
\includegraphics[width= 0.85 \linewidth]{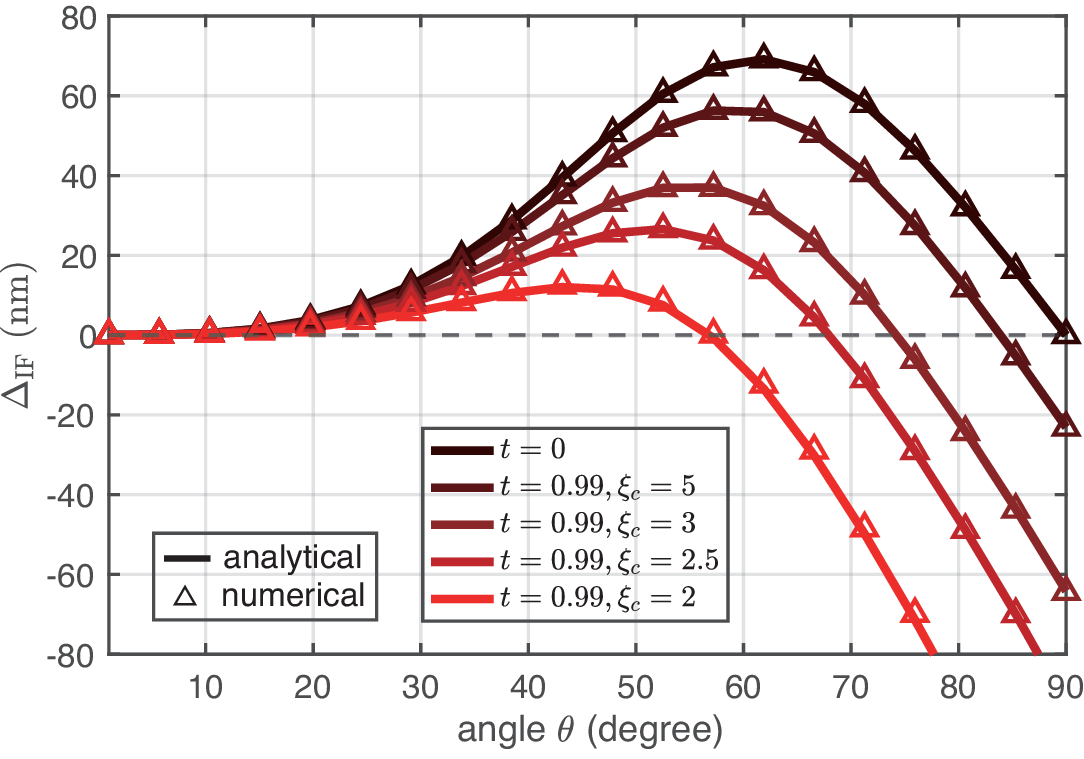}
\caption{Analytical (solid curves) and numerical (triangles) results for the spatial IF shifts of a circularly polarized TGSM beam under partial reflection with $t=0.99$ and variable $\xi_c$. The black curve shows the result for $t=0$ (spatial IF shift is purely induced by the spin Hall effect). The simulation parameters are: $n_1 = 1$, $n_2 = 1.515$, $a_x = 1$, $a_y = \mathrm{e}^{\mathrm{i} \pi/2}$, $\lambda_0 = 633$~nm, $\sigma_I = 1$~mm.}
\label{fig3}
\end{figure}

Next, we consider the case of partial reflection of a circularly polarized TGSM beam ensemble. To this end, we let $a_x = 1$ and $a_y = \mathrm{e}^{\mathrm{i} \pi/2}$. In this case, the GH shift vanishes identically (see End Matter) and the IF shift consists of two contributions: one from the spin Hall effect and the other from the coherence Hall effect, which makes it possible to cancel the overall IF shift altogether by tuning the sign of $t$ at sufficiently high coherence level. We illustrate this novel effect in Fig.~\ref{fig3}.  The curve for $t = 0$ shows the spin-induced IF shift, which is always positive in the angular interval $\theta \in (0^\circ, 90^\circ)$ for $a_x = 1$ and $a_y = \mathrm{e}^{\mathrm{i} \pi/2}$, and becomes negative for $a_x = 1$ and $a_y = \mathrm{e}^{-\mathrm{i} \pi/2}$. Introducing a twist phase with $t = 0.99$, we observe the cancellation of the IF shift at a particular angle of incidence. As $\xi_c$ decreases, the cancellation angle gradually shifts to smaller values. The analytical and numerical results in Fig.~\ref{fig3} are in excellent agreement.

\begin{figure}[t]
\centering
\includegraphics[width = \linewidth]{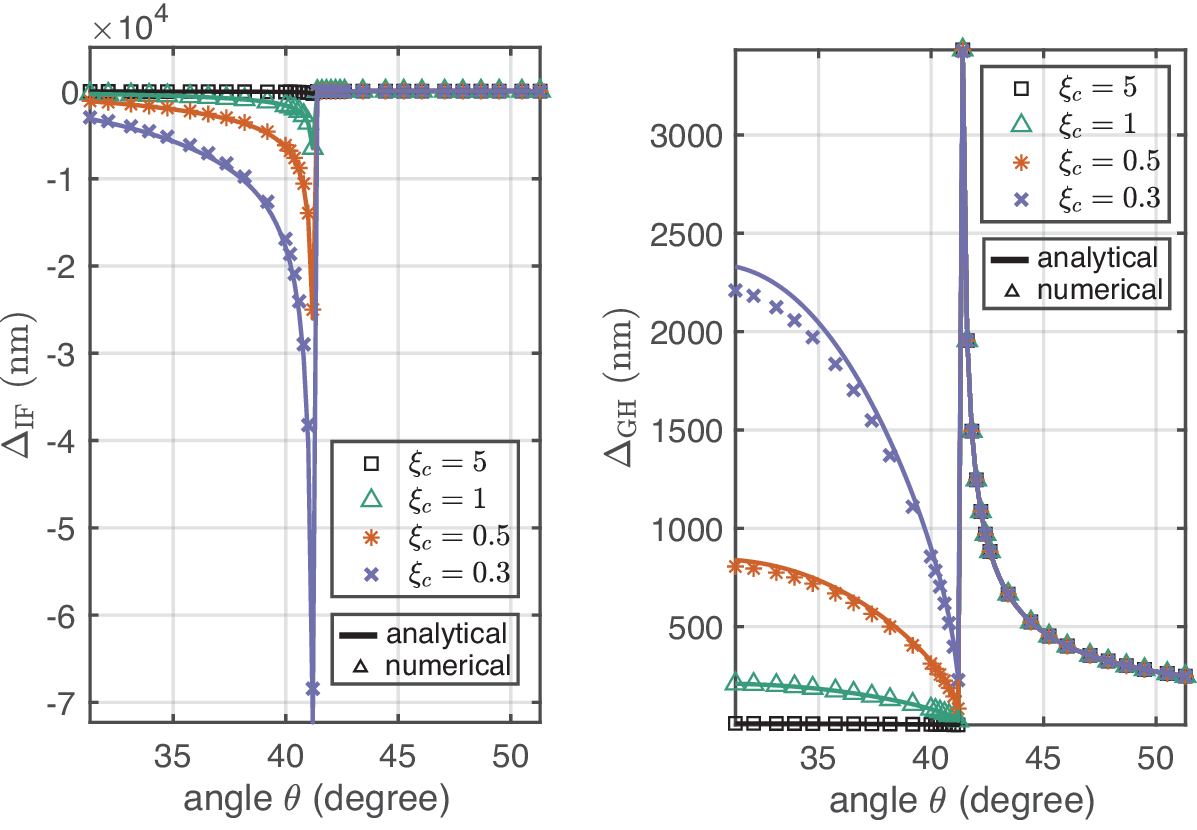}
\caption{Analytical (curves) and numerical (markers) results for the GH and IF shifts of a linearly polarized TGSM beam under reflection with $t = 0.99$ and varying $\xi_c$. The simulation parameters are: $n_1 = 1.515$, $n_2 = 1$, $a_x = a_y = 1$, $\lambda_0 = 633$~nm, $\sigma_I = 1$~mm. The beam undergoes total internal reflection for $\theta \geq 41.305^\circ$ and partial reflection for $\theta < 41.305^\circ$.}
\label{fig4}
\end{figure}

Finally, we consider total internal reflection of TGSM beams from the glass/air interface. The refractive indices of the two media are $n_1 = 1.515$ and $n_2 = 1$, giving the critical angle for total internal reflection as $\theta_c = \mathrm{arcsin}(n_2 / n_1) = 41.305^\circ$. Our analysis, see End Matter, indicates that above the critical angle $\bm{\Lambda}=\bm{\Theta}=0$ which implies that for the incidence angles $\theta\geq \theta_c$, neither spatial nor angular shifts depend on either source coherence or its phase-space structure. We illustrate this point in Fig.~\ref{fig4} by comparing the GH and IF shifts with $t=0.99$ and variable $\xi_c$ for incidence angles greater and smaller than $\theta_c$. 

In conclusion, we have discovered a link between the phase-space nonseparability of an incident paraxial wave packet and spatial shifts of the packet in reflection from an interface separating two transparent media. We have derived a general expression for GH and IF shifts of a phase-space nonseparable wave packet of any degree of spatial coherence and illustrated our results with a TGSM beam reflection from the interface. Our results elucidate the role of spatial coherence in wave packet reflection from the interface and resolve a long-standing controversy about the role of spatial coherence of the source in spatial GH shift formation. Instructively and contrary to the common wisdom, we have shown that the most pronounced GH and IF shifts occur in partial reflection of phase-space nonseparable wave packets. This circumstance and the giant enhancement of the shifts in the low-coherence limit can facilitate their observation in the laboratory. The coherence GH and coherence Hall effects, which we have discovered, reveal subtle and, to our knowledge, hitherto unnoticed fundamental aspects of the interaction of structured wave packets with matter. 

The authors acknowledge financial support from the National Natural Science Foundation of China (12274310) and the Natural Sciences and Engineering Research Council of Canada (RGPIN-2025-04064).


\begin{thebibliography}{99}

\bibitem{goos1947neuer}
F. Goos and H. Hänchen, Ein neuer und fundamentaler versuch zur totalreflexion. Ann. Phys. \textbf{436}, 333 (1947).

\bibitem{fedorov1955k}
F. I. Fedorov, K teorii polnogo otrazheniya, Dokl. Akad. Nauk SSSR \textbf{105}, 465 (1955).

\bibitem{Imbert72}
C. Imbert, Calculation and experimental proof of the transverse shift induced by total internal reflection of a circularly polarized light beam, Phys. Rev. D \textbf{5}, 787 (1972).

\bibitem{Ra73}
J. W. Ra, H. Bertoni, and L. Felsen, Reflection and transmission of beams at a dielectric interface, SIAM J. Appl. Math. \textbf{24}, 396 (1973).

\bibitem{Merano09}
M. Merano, A. Aiello, M. P. van Exter, and J. P. Woerdman, Observing angular deviations in the specular reflection of a light beam, Nat. Photonics \textbf{3}, 337 (2009).

\bibitem{ignatovich2004neutron}
V. Ignatovich, Neutron reflection from condensed matter, the Goos-Hänchen effect and coherence, Phys. Lett. A \textbf{322}, 36 (2004).

\bibitem{de2010observation}
V.-O. de Haan, J. Plomp, T. M. Rekveldt, W. H. Kraan, A. A. van Well, R. M. Dalgliesh, and S. Langridge, Observation of the Goos-Hänchen shift with neutrons, Phys. Rev. Lett. \textbf{104}, 010401 (2010).

\bibitem{tamasaku2002goos}
K. Tamasaku and T. Ishikawa, The Goos-Hänchen effect at Bragg diffraction, Acta Cryst. \textbf{A58}, 408 (2002). 

\bibitem{beenakker2009quantum}
C. Beenakker, R. Sepkhanov, A. Akhmerov, and J. Tworzydlo, Quantum Goos-Hänchen effect in graphene, Phys. Rev. Lett. \textbf{102}, 146804 (2009).

\bibitem{wu2011valley}
Z. Wu, F. Zhai, F. Peeters, H. Xu, and K. Chang, Valley-dependent Brewster angles and Goos-Hänchen effect in strained graphene, Phys. Rev. Lett. \textbf{106}, 176802 (2011).

\bibitem{dadoenkova2012huge}
Y. S. Dadoenkova, N. N. Dadoenkova, I. L. Lyubchanskii, M. L. Sokolovskyy, J. W. Kłos, J. Romero-Vivas, and M. Krawczyk, Huge Goos-Hänchen effect for spin waves: A promising tool for study magnetic properties at interfaces, Appl. Phys. Lett. \textbf{101}, 042404 (2012).

\bibitem{huang2008goos}
J. Huang, Z. Duan, H. Y. Ling, and W. Zhang, Goos-Hänchen-like shifts in atom optics, Phys. Rev. A \textbf{77}, 063608 (2008).

\bibitem{mckay2025observation}
S. McKay, V. O. de Haan, J. Leiner, S. R. Parnell, R. M. Dalgliesh, P. Boeni, L. J. Bannenberg, Q. Le Thien, D. V. Baxter, G. Ortiz, and R. Pynn, Observation of a giant Goos-Hänchen shift for matter waves, Phys. Rev. Lett. \textbf{134}, 093803 (2025). 

\bibitem{ponomarenko2022goos}
S. A. Ponomarenko, J. Zhang, and G. P. Agrawal, Goos-Hänchen shift at a temporal boundary, Phys. Rev. A \textbf{106}, L061501 (2022).

\bibitem{Onoda04}
M. Onoda, S. Murakami, and N. Nagaosa, Hall effect of light, Phys. Rev. Lett. \textbf{93}, 083901 (2004).

\bibitem{Bliokh06}
K. Y. Bliokh and Y. P. Bliokh, Conservation of angular momentum, transverse shift, and spin Hall effect in reflection and refraction of an electromagnetic wave packet, Phys. Rev. Lett. \textbf{96}, 073903 (2006).

\bibitem{Hosten08}
O. Hosten and P. Kwiat, Observation of the spin Hall effect of light via weak measurements, Science \textbf{319}, 787 (2008).

\bibitem{bliokh2013goos}
K. Y. Bliokh and A. Aiello, Goos--Hänchen and Imbert--Fedorov beam shifts: an overview, J. Opt. \textbf{15}, 014001 (2013).

\bibitem{Kim23}
M. Kim, Y. Yang, D. Lee, Y. Kim, H. Kim, and J. Rho, Spin Hall effect of light: From fundamentals to recent advancements, Laser Photonics Rev. \textbf{17}, 2200046 (2023).

\bibitem{zhou2021temperature}
X. Zhou, P. Tang, C. Yang, S. Liu, and Z. Luo, Temperature-dependent Goos-Hänchen shifts in a symmetrical graphene-cladding waveguide, Results Phys. \textbf{24}, 104100 (2021).

\bibitem{wang2015optical}
X. Wang, M. Sang, W. Yuan, Y. Nie, and H. Luo, Optical relative humidity sensing based on oscillating wave-enhanced Goos-Hänchen shift, IEEE Photonics Technol. Lett. \textbf{28}, 264 (2016).

\bibitem{Zhu19Generalized}
T. Zhu, Y. Lou, Y. Zhou, J. Zhang, J. Huang, Y. Li, H. Luo, S. Wen, S. Zhu, Q. Gong, M. Qiu, and Z. Ruan, Generalized spatial diﬀerentiation from the spin Hall eﬀect of light and its application in image processing of edge detection, Phys. Rev. Applied \textbf{11}, 034043 (2019).

\bibitem{Wang22Photonic}
R. Wang, S. He, and H. Luo, Photonic spin-Hall differential microscopy, Phys. Rev. Applied \textbf{18}, 044016 (2022).

\bibitem{Liu24PRL}
J. Liu, Q. Yang, Y. Shou, S. Chen, W. Shu, G. Chen, S. Wen, and H. Luo, Metasurface-assisted quantum nonlocal weak-measurement microscopy, Phys. Rev. Lett. \textbf{132}, 043601 (2024).

\bibitem{wu2019giant}
F. Wu, J. Wu, Z. Guo, H. Jiang, Y. Sun, Y. Li, J. Ren, and H. Chen, Giant enhancement of the Goos-Hänchen shift assisted by quasibound states in the continuum, Phys. Rev. Applied \textbf{12}, 014028 (2019).

\bibitem{wu2021observation}
J. Wu, X. Xu, X. Su, S. Zhao, C. Wu, Y. Sun, Y. Li, F. Wu, Z. Guo, H. Jiang, and H. Chen, Observation of giant extrinsic chirality empowered by quasi-bound states in the continuum, Phys. Rev. Applied \textbf{16}, 064018 (2021).

\bibitem{frank2014goos}
A. I. Frank, On the Goos-H\"{a}nchen effect in neutron optics, J. Phys. Conf. Ser. \textbf{528}, 012029 (2014).

\bibitem{fedoseyev2001spin}
V. Fedoseyev, Spin-independent transverse shift of the centre of gravity of a reflected and of a refracted light beam, Opt. Commun. \textbf{193}, 9 (2001).

\bibitem{dasgupta2006experimental}
R. Dasgupta and P. Gupta, Experimental observation of spin-independent transverse shift of the centre of gravity of a reflected Laguerre--Gaussian light beam, Opt. Commun. \textbf{257}, 91 (2006).

\bibitem{okuda2008significant}
H. Okuda and H. Sasada, Significant deformations and propagation variations of Laguerre--Gaussian beams reflected and transmitted at a dielectric interface, J. Opt. Soc. Am. A \textbf{25}, 881 (2008).

\bibitem{bliokh2009goos}
K. Y. Bliokh, I. V. Shadrivov, and Y. S. Kivshar, Goos--Hänchen and Imbert--Fedorov shifts of polarized vortex beams, Opt. Lett. \textbf{34}, 389 (2009).

\bibitem{bekshaev2009oblique}
A. Y. Bekshaev, Oblique section of a paraxial light beam: criteria for azimuthal energy flow and orbital angular momentum, J. Opt. A: Pure Appl. Opt. \textbf{11}, 094003 (2009).

\bibitem{merano2010orbital}
M. Merano, N. Hermosa, J. Woerdman, and A. Aiello, How orbital angular momentum affects beam shifts in optical reflection, Phys. Rev. A \textbf{82}, 023817 (2010).

\bibitem{Dennis12}
M. R. Dennis and J. B. Götte, Topological aberration of optical vortex beams: Determining dielectric interfaces by optical singularity shifts, Phys. Rev. Lett. \textbf{109}, 183903 (2012). 

\bibitem{Loffler12}
W. Löffler, A. Aiello, and J. P. Woerdman, Observation of orbital angular momentum sidebands due to optical reflection, Phys. Rev. Lett. \textbf{109}, 113602 (2012). 

\bibitem{Barros24}
R. F. Barros, S. Bej, M. Hiekkamäki, M. Ornigotti, and R. Fickler, Observation of the topological aberrations of twisted light, Nat. Commun. \textbf{15}, 8162 (2024).


\bibitem{le2025entangled}
Q. Le Thien, R. Pynn, and G. Ortiz, Entangled-beam reflectometry and Goos-Hänchen shift, Phys. Rev. Lett. \textbf{134}, 093802 (2025).

\bibitem{simon1989nonspecular}
R. Simon and T. Tamir, Nonspecular phenomena in partly coherent beams reflected by multilayered structures, J. Opt. Soc. Am. A \textbf{6}, 18 (1989).

\bibitem{aiello2011role}
A. Aiello and J. Woerdman, Role of spatial coherence in Goos-Hänchen and Imbert-Fedorov shifts, Opt. Lett. \textbf{36}, 3151 (2011).

\bibitem{tang2025effect}
M. Tang, J. Laatikainen, M. Ornigotti, T. Setälä, and A. Norrman, Effect of partial polarization on Goos--Hänchen and Imbert--Fedorov shifts, Opt. Lett. \textbf{50}, 447 (2025).

\bibitem{loffler2012spatial}
W. Löffler, A. Aiello, and J. P. Woerdman, Spatial coherence and optical beam shifts, Phys. Rev. Lett. \textbf{109}, 213901 (2012).

\bibitem{merano2012observation}
M. Merano, G. Umbriaco, and G. Mistura, Observation of nonspecular effects for Gaussian Schell-model light beams, Phys. Rev. A \textbf{86}, 033842 (2012).

\bibitem{wang2008influence}
L.-Q. Wang, L.-G. Wang, S.-Y. Zhu, and M. S. Zubairy, The influence of spatial coherence on the Goos--Hänchen shift at total internal reflection, J. Phys. B \textbf{41}, 055401 (2008).

\bibitem{wang2013goos}
L.-G. Wang, S.-Y. Zhu, and M. S. Zubairy, Goos-Hänchen shifts of partially coherent light fields, Phys. Rev. Lett. \textbf{111}, 223901 (2013).

\bibitem{Simon93}
R. Simon and N. Mukunda, Twisted Gaussian Schell-model beams, J. Opt. Soc. Am. A \textbf{10}, 2008 (1993).

\bibitem{sundar1993twisted}
K. Sundar, R. Simon, and N. Mukunda, Twisted Gaussian Schell-model beams. II. Spectrum analysis and propagation characteristics, J. Opt. Soc. Am. A \textbf{10}, 2017 (1993).

\bibitem{ponomarenko2021twist}
S. A. Ponomarenko, Twist phase and classical entanglement of partially coherent light, Opt. Lett. \textbf{46}, 5958 (2021).

\bibitem{mandel1995optical}
L. Mandel and E. Wolf, \textit{Optical Coherence and Quantum Optics} (Cambridge University Press, 1995).

\bibitem{schleich2015quantum}
W. P. Schleich, \textit{Quantum Optics in Phase Space} (John Wiley \& Sons, 2015). 

\bibitem {SM}
See Supplemental Material which includes Ref.~\cite{novotny2012,jackson2021,gori2015,gradshteyn2014,arfken2011}, derivations of a general analytical expression for the beam shifts in the Wigner picture, the analytical expression for TGSM beam centroid position, the limit of a GH shift of a polarized, coherent, phase-separable beam under total internal reflection, and coherent-mode representation of a TGSM source, as well as a detailed description of our numerical procedure. 

\bibitem{novotny2012}
L. Novotny and B. Hecht, \textit{Principles of Nano-Optics} (Cambridge University Press, 2012), 2nd ed.

\bibitem{jackson2021}
J. D. Jackson, \textit{Classical Electrodynamics} (John Wiley \& Sons, 2021).

\bibitem{gori2015}
F. Gori and M. Santarsiero, Twisted Gaussian Schell-model beams as series of partially coherent modified Bessel–Gauss beams,
Opt. Lett., {\bf 40} 1587 (2015).

\bibitem{gradshteyn2014}
I. S. Gradshteyn and I. M. Ryzhik, \textit{Table of Integrals, Series, and Products} (Academic Press, 2014).

\bibitem{arfken2011}
G. B. Arfken, H. J. Weber, and F. E. Harris, \textit{Mathematical Methods for Physicists: A Comprehensive Guide} (Academic Press, 2011).


\bibitem{Ari94}
A. T. Friberg, E. Tervonen, and J. Turunen, Interpretation and experimental demonstration of twisted Gaussian Schell-model beams, J. Opt. Soc. Am. A {\bf 11}, 1818 (1994).

\bibitem{Wang19}
H. Wang, X. Peng, L. Liu, F. Wang. Y. Cai, and S. A. Ponomarenko, Generating bona fide twisted Gaussian Schell-model beams, Opt. Lett. {\bf 44}, 3709 (2019). 

\bibitem{Wang25}
H. Wang, Y. Wang, H. Peng, L. Liu, Y. Cai, and F. Wang, Real-time synthesis of twisted Gaussian Schell-model beams and their applications in suppressing the turbulence-induced scintillation and beam wander, Opt. Lett. {\bf 50}, 4342 (2025).

\bibitem{PSA01}
S. A. Ponomarenko, Twisted Gaussian Schell-model solitons, Phys. Rev. E {\bf 64}, 036618 (2001). 

\bibitem{Yahong25}
B. Li, Y. Chen, W. Deng, T. Wang, L. Wan, and T. Yu, Spin--orbit interactions of twisted random light, APL Photon. {\bf 10}, 046119 (2025).

\bibitem{Cai06}
Y. Cai and S. He, Propagation of a partially coherent twisted anisotropic Gaussian Schell-model beam in a turbulent atmosphere, Appl. Phys. Lett. {\bf 89}, 041117 (2006).

\bibitem{Wang12}
F. Wang. Y. Cai,  H. T. Eyyuboglu, and Y. Baykal, Twist phase-induced reduction in scintillation of a partially coherent beam in turbulent atmosphere, Opt. Lett. {\bf 37}, 184 (2012). 

\bibitem{Olga09}
Y. Cai, Q. Lin, and O. Korotkova, Ghost imaging with twisted Gaussian Schell-model beam, Opt. Express {\bf 17}, 2453 (2009). 

\bibitem{Olga12}
Z. Tong and O. Korotkova, Beyond the classical Rayleigh limit with twisted light, Opt. Lett. {\bf 37}, 2595 (2012).

\bibitem{Hutter20}
L. Hutter, G. Lima, and S. P. Walborn, Boosting entanglement generation in down-conversion with incoherent illumination, Phys. Rev. Lett. \textbf{125}, 193602 (2020). 

\bibitem{gbur16}
G. J.  Gbur, \textit{Singular Optics} (CRC Press, 2016).

\bibitem{Wang18recent}
X. Wang, Z. Nie, Y. Liang, J. Wang, T. Li, and B. Jia, Recent advances on optical vortex generation, Nanophotonics \textbf{7}, 1533 (2018).

\end{thebibliography}

\section{End Matter}
\renewcommand{\theequation}{A\arabic{equation}}
\setcounter{equation}{0}

The (scalar) WDF of a TGSM wave packet is given by the expression~\cite{ponomarenko2021twist}
\begin{equation}\label{WDF-A}
    \mathcal{W}_i (\Bkappa,\BR)= \mathcal{W}_-(\kappa_x,Y)\mathcal{W}_+(\kappa_y,X),
\end{equation}
where 
\begin{subequations}\label{Wig-TGSM}
    \begin{eqnarray}
        \mathcal{W}_-(\kappa_x,Y)\propto e^{-Y^2/2\sigma_I^2}e^{-(\kappa_x-uY)^2 \sigma_{\mathrm{eff}}^2/2}, \\
         \mathcal{W}_+(\kappa_y,X)\propto e^{-X^2/2\sigma_I^2}e^{-(\kappa_y+uX)^2 \sigma_{\mathrm{eff}}^2/2},
        \end{eqnarray}
    \end{subequations}
and we dropped irrelevant normalization factors. Here $\sigma_I$ and $\sigma_c$ are the rms beam and coherence widths and 
\begin{equation}\label{sigef}
    \frac{1}{\sigma_{\mathrm{eff}}^2}=\frac{1}{\sigma_c^2}+\frac{1}{4\sigma_I^2}.
\end{equation}
We conclude from Eqs.~(\ref{Wig-TGSM}) that the momentum of the wave packet in the $x$ ($y$) direction is (classically) entangled with the corresponding $y$ ($x$) coordinate. The degree of entanglement is quantified by the strength of the twist parameter $u$~\cite{ponomarenko2021twist} which obeys the constraint;
\begin{equation}\label{u}
    -1/\sigma_c^2\leq u\leq 1/\sigma_c^2,
\end{equation}
and hence vanishes in the fully coherent limit, $\sigma_c\rightarrow\infty$. 

The reflection-amplitude dependent coefficients for spatial and angular GH and IF shifts, cf.~Eq.~(\ref{Rc}) in the main text, read
\begin{equation}\label{Del}
\begin{split}
    \bm{\Delta}&= - \frac{|a_x|^2 |r_p|^2 \mathrm{Im}X_p+|a_y|^2 |r_s|^2 \mathrm{Im}X_s}{|a_x|^2 |r_p|^2+|a_y|^2|r_s|^2}\Be_x' \\
&+\frac{|r_p|^2 \mathrm{Im}(Y_p a_x^* a_y) - |r_s|^2 \mathrm{Im}(Y_s a_x a_y^*)}{|a_x|^2 |r_p|^2+|a_y|^2|r_s|^2}\Be_y',
\end{split}
\end{equation}
\begin{equation}
\begin{split}\label{Lda}
    \bm{\Lambda}&=\frac{ |r_p|^2\mathrm{Re}(Y_pa_x^*a_y)-|r_s|^2 \mathrm{Re}(Y_sa_xa_y^*)}{|a_x|^2r_{p}|^2+|a_y|^2|r_s|^2}\Be_x' \\
&+\frac{|a_x|^2|r_p|^2 \mathrm{Re}X_p+|a_y|^2|r_s|^2 \mathrm{Re}X_s}{|a_x|^2|r_{p}|^2+|a_y|^2|r_s|^2}\Be_y',
\end{split}
\end{equation}
and
\begin{equation}\label{Tta}
\begin{split}
    \bm{\Theta}&=\frac{|a_x|^2|r_p|^2 \mathrm{Re}X_p+|a_y|^2|r_s|^2 \mathrm{Re}X_s}{|a_x|^2|r_p|^2+|a_y|^2|r_s|^2}\Be_x' \\
    &+\frac{|r_s|^2\mathrm{Re}(Y_s a_x a_y^*)-|r_p|^2\mathrm{Re}(Y_p a_x^*a_y)}{|a_x|^2|r_p|^2+|a_y|^2|r_s|^2}\Be_y'.
\end{split}
\end{equation}
Here $\Be_x'$ and $\Be_y'$ are the unit vectors along the $x'$ and $y'$ directions, respectively and we introduce the notation
\begin{gather}\label{XY}
    X_{p,s} =\frac{1}{k}\frac{\partial \ln r_{p,s}}{\partial\theta}, \quad Y_{p,s}=\frac{1}{k}\left(1+\frac{r_{s,p}}{r_{p,s}}\right)\cot\theta.
\end{gather}

{\it Partial reflection.---}In this case, $\mathrm{Im}r_s=\mathrm{Im} r_p=0$ (for transparent media), implying that $\mathrm{Im}X_{p,s}=0$ and $\mathrm{Im}Y_{p,s}=0$. It follows from Eq.~(\ref{Del}) that the $\Be'_x$ component of $\bm{\Delta}$ vanishes, $\Delta_{x'}=0$. Further, for circular polarizations, $a_x=1$, $a_y=e^{\pm \mathrm{i} \pi/2}$, say, we can infer from Eq.~(\ref{Lda}) that the $\Be'_x$ component of $\bm{\Lambda}$ equals to zero as well, $\Lambda_{x'}=0$. Therefore, the spatial GH shift vanishes identically.

{\it Total internal reflection.---}In this case, $r_s$ and $r_p$ are unimodular complex functions, and $X_s$ as well as $X_p$ are purely imaginary. Therefore, the $\Be'_y$ component of $\bm{\Lambda}$ in Eq.~(\ref{Lda}) and the $\Be'_x$ component of $\bm{\Theta}$ in Eq.~(\ref{Tta}) vanish. To prove that the $\Be'_x$ component of $\bm{\Lambda}$ and the $\Be'_y$ component of $\bm{\Theta}$ also vanish, we proceed as follows:
\begin{align}
Y_s a_x a_y^\ast & = \frac{1}{k}  \mathrm{cot} \theta |a_x||a_y|  \left(1+  e^{\mathrm{i} \Delta \phi} \right) e^{\mathrm{i} \Delta \chi}, \\
Y_p a_x^\ast a_y & =  \frac{1}{k}  \mathrm{cot} \theta |a_x||a_y| \left(1+  e^{-\mathrm{i} \Delta \phi}  \right) e^{-\mathrm{i} \Delta \chi},
\end{align}
where $\Delta \phi = \mathrm{arg}(r_p) - \mathrm{arg}(r_s)$ and $\Delta \chi = \mathrm{arg}(a_x) - \mathrm{arg}(a_y)$. Thus, recalling that $|r_s|^2=|r_p|^2=1$, we conclude that
\begin{equation}\label{zero}
 |r_s|^2 \mathrm{Re}(Y_s a_x a_y^\ast) - |r_p|^2 \mathrm{Re}(Y_p a_x^\ast a_y) = 0.
\end{equation}
Combining Eqs.~(\ref{zero}),~(\ref{Lda}), and (\ref{Tta}), we obtain 
\begin{align}
\label{Lambda70TIR}
    \bm{\Lambda} & = 0, \\
\label{Theta79TIR}
    \bm{\Theta} & = 0.
\end{align}

\end{document}